\documentclass{aa}
\usepackage{natbib}
\bibpunct{(}{)}{;}{a}{}{,}
\usepackage{graphicx}
\newcommand{\curl}{\mathbf{curl}}
\newcommand{\cwl}{\ensuremath{\raisebox{0.6ex}{--}\hspace{-0.55em}\lambda_\mathrm{e}}}
\newcommand{\href}[2]{#2}

\begin{document}

\title{Efficient acceleration and radiation in Poynting flux powered
  GRB outflows}

\author{G. Drenkhahn\thanks{e-mail:
    \texttt{georg@mpa-garching.mpg.de}} \and H. C. Spruit}

\institute{Max-Planck-Institut f\"ur Astrophysik, Postfach~1317,
  85741~Garching bei M\"unchen, Germany}

\date{Received 20 February 2002 / Accepted 4 June 2002}

\abstract{We investigate the effects of magnetic energy release by
  local magnetic dissipation processes in Poynting flux-powered GRBs.
  For typical GRB parameters (energy and baryon loading) the
  dissipation takes place mainly outside the photosphere, producing
  non-thermal radiation.  This process converts the total burst energy
  into prompt radiation at an efficiency of 10--50\%.  At the same
  time the dissipation has the effect of accelerating the flow to a
  large Lorentz factor.  For higher baryon loading, the dissipation
  takes place mostly inside the photosphere, the efficiency of
  conversion of magnetic energy into radiation is lower, and an X-ray
  flash results instead of a GRB.  We demonstrate these effects with
  numerical one-dimensional steady relativistic MHD calculations.
  
  \keywords{Gamma rays: bursts -- Magnetic fields --
    Magnetohydrodynamics (MHD) -- Stars: winds, outflows}}

\maketitle

\defcitealias{spruit:01}{Paper~I}
\defcitealias{drenkhahn:02a}{Paper~II}

\section{Introduction}

High luminosity outflows from $\gamma$-ray bursts (GRBs) must have
large Lorentz factor to overcome the compactness problem
\citep[e.g.][]{piran:99}.  To fulfil this requirement the mass loading
can only be small so that the total energy density exceeds greatly the
rest mass energy density.  Poynting flux can carry large energy
amounts though vacuum which provides a mechanism to transport energy
without the need of matter.  The release of electromagnetic energy by
the central engine of a GRB is part of many models.  E.g.\ tori in
merger scenarios may by highly magnetised due to the field
amplification by the differential rotation
\citep{narayan:92,thompson:94,meszaros:97,katz:97}.  Alternative
models involve highly magnetised millisecond pulsars
\citep{usov:92,spruit:99}.  In all cases the rotational energy of a
compact object will be tapped and the rotating magnetic field produces
a Poynting flux.

While Poynting flux is thus a plausible way of powering a GRB, it is
not a priori clear how this energy flux is converted into the observed
$\gamma$-rays.  To accelerate the matter to the observed high Lorentz
factors, a part of the Poynting flux must be converted into kinetic
energy.  This energy later powers the afterglow when it is released in
an external shock.  Since the prompt emission in most cases accounts
for the bulk of the observed radiation, a mechanism is needed to
efficiently convert a magnetic energy flux into non-thermal radiation.

For the acceleration of the flow one can think of magnetocentrifugal
effects.  But trying to explaining the flow acceleration by stationary
ideal MHD processes is problematic.  For the purely radial
magnetically driven stellar wind \citep[modelled originally
by][]{weber:67,mestel:68} the radial gradient of the magnetic pressure
and the inward pointing magnetic tension force act against each other.
Especially in the case where the flow is relativistic from the
beginning there is a balance between these forces so that such purely
radial flows are not accelerating.  However, an acceleration takes
place if the flow lines diverges faster with radius than in the radial
case beyond the fast critical point.  The tension and pressure
gradient forces are out of balance and Poynting flux to kinetic energy
flux conversion occurs \citep{begelman:94,daigne:01}.  A magnetic
acceleration model which uses only ideal MHD thus has to provide just
the right flow divergence for acceleration to take place.  As we show
in this paper, a better alternative is a flow in which part of the
magnetic energy density is dissipated locally.  The decrease of
magnetic energy density with distance in such a model causes effective
acceleration \citep{lyubarsky:01,drenkhahn:02a}, while at the same
time providing an efficient energy source for the observed
$\gamma$-rays.

The currently most accepted model explaining the high energy prompt
emission of GRBs is the internal shock model
\citep{rees:92,rees:94,sari:97}.  Variations of the central engine
luminosity produces flow shells with different Lorentz factors which
collide.  Through these collisions a part of the kinetic energy is
transfered into prompt radiation.  The energy conversion is only
efficient if the spread in Lorentz factors is large
\citep{kumar:99,panaitescu:99,beloborodov:00,kobayashi:01}.  The
observed ratio of afterglow and prompt emission indicates a high
efficiency of energy conversion into prompt emission.  While this
observation does not rule out the internal shock model, it does put
strong constraints on it.

If one allows for non-ideal MHD processes magnetic energy can be
transfered to the matter by dissipation through reconnection.  For
this to happen, there must be small scale variations in the magnetic
field.  The energy that is released by washing out these variations
can be converted into radiation.  We call this the `free magnetic
energy' in the flow.  An example of such small scale variations would
be the `striped' wind \citep{coroniti:90} that results from the
rotation of an inclined magnetic dipole.  The distance between
neighbouring stripes of different field direction in this case is $\pi
v/\Omega$, where $v$ is the outflow speed and $\Omega$ is the dipole's
angular frequency.  This scale is of the order of the size of the
central engine (assumed to be a relativistic object).  In general, all
non-axisymmetric components of the magnetic field of the central
engine produce such variations.  If the magnetic configuration is
predominantly non-axisymmetric, almost all of the Poynting flux is in
the form of a field that changes sign on such a small length scale.
This is the model we will use for the quantitative calculations below.
We note, however, that even an axisymmetric rotating field can in
principle produce small length scales.  The outflow near the axis of
an axisymmetric MHD flow is spiral-like.  This configuration is kink
unstable so that field components can reconnect across the rotation
axis.  For a discussion of this point see \citet{spruit:01}, hereafter
\citetalias{spruit:01}.  As we showed in \citet{drenkhahn:02a}
(\citetalias{drenkhahn:02a} from here on), such kink-produced
irregularities are somewhat less efficient at converting magnetic
energy.  Since perfect axisymmetry is a special case we regard the
non-axisymmetric case in this study.

Fast reconnection leads to a decay of the magnetic field.  The flow
accelerates since the field decay induces an additional outward
gradient in the magnetic pressure.  In \citetalias{drenkhahn:02a} we
explored the dynamical effects of the Poynting flux dissipation in the
outflow.  We found there that for fiducial GRB parameters a large
amount of the Poynting flux energy is converted to kinetic energy.
Also, a great part of the dissipation happens in the optically thin
region so that a potentially large fraction could be converted into
non-thermal, prompt radiation.  This model offers an alternative to
the internal shock model in explaining the prompt emission by local
dissipation of free magnetic energy.

The results presented in \citetalias{drenkhahn:02a} were based on an
analytic approximation for the flow.  In the present paper, we relax
this approximation, and analyse a Poynting flux powered wind
numerically.  The results confirm the main results from the analytical
study, but in addition allow us to determine which fractions of the
Poynting flux are converted into thermal and non-thermal radiation.
In this way, we can also determine the conditions under which a true
GRB, as opposed to an `X-ray flash' \citep{heise:01,heise:02} is
produced by a magnetically powered outflow.

\section{The Model}

We consider a radial outflow of magnetised plasma with the magnetic
field being aligned transversal to the flow direction.  The field
contains small scale variations in direction from which energy is
released.  We parameterise the variation by introducing a
\emph{variation length scale} $\lambda\approx 2\pi c/\Omega$ on which
the direction of the field changes.  Field variations in the outflow
are naturally produced by any non-axisymmetric component of a rotating
magnetic field.

The rate of magnetic energy dissipation is governed by the
reconnection rate between neighbouring regions of different field line
direction.  For highly symmetric initial conditions, the initial
reconnection process is sensitive to the microscopic diffusion rate,
but this situation is rarely relevant in astrophysics.  Instead, the
field reconnects by `rapid reconnection' processes, in which the
reconnection speed depends only logarithmically on the microscopic
transport coefficients \citep{petschek:64,parker:79,priest:00}.  Since
the reconnection rate is an important factor influencing the results,
we keep track of its effect by a parameter study.  For this purpose we
write the reconnection time scale $\tau_\mathrm{co}$ across the
variation length scale $\lambda_\mathrm{co}$ as $\tau_\mathrm{co} =
\lambda_\mathrm{co}/v_r$, where the reconnection speed $v_r$ is the
velocity at which field lines of different directions are brought
together by the dynamics of the reconnection process.  We regard this
process in the comoving frame moving with the bulk large-scale flow.
The speed $v_r$ is known to scale with the Alfv\'en speed $v_A$, i.e.
$v_r = \epsilon v_A$ where $\epsilon$ is a numerical factor $<1$.  For
rapid reconnection in 2 dimensions, for example, numerical results
\citep[e.g.][]{biskamp:86} show that $\epsilon$ can be of the order
$0.1$.  Since we do not know the density of reconnection centres the
overall rate of field dissipation is still unknown.  By adjusting the
parameter $\epsilon$ towards lower values we compensate for this
ignorance.  Thus, we take $\epsilon$ as a measure for both, the
reconnection speed and the density of reconnection centres.

As a second parameter of less importance, we introduce the fraction
$\mu^2$ of the magnetic energy density that cannot be dissipated by
local reconnection.  If the field is the result of the winding-up of a
completely non-axisymmetric field (i.e.\ with vanishing azimuthal
average), the direction of the field lines changes completely over one
variation length, and we have $\mu=0$.  If, on the other hand, the
axisymmetric component does not vanish, we have $\mu =
B_{\Uparrow}/B$, where $B_{\Uparrow}$ and $B$ are the amplitudes of
the axisymmetric component and the total field. In most of our study,
we will assume that the Poynting flux decays completely which
correspond to $\mu=0$.

The released magnetic energy is converted into thermal and kinetic
energy.  The direct conversion into kinetic energy is possible because
the field dissipation induces an additional outward gradient in the
magnetic pressure.  So even if there is no thermal pressure gradient
(as in the cold approximation used in \citetalias{drenkhahn:02a}) part
of the released energy accelerates the flow directly.

Though small scale structures in the flow are an essential part of the
model, we will only work with quantities which are averaged over these
small scales.  One can consider the flow to be stationary on large
length scales so that we can do a time independent calculation to
obtain general results.  The adjustable variable $\epsilon$ accounts
for the unknown processes on small scales and includes effects
introduced by the averaging.

The dissipated magnetic energy initially takes the form of internal
(thermal) energy of the gas.  If the cooling time is long compared
with the expansion time scale of the flow, the energy is mostly
converted to kinetic energy, through adiabatic expansion.  This is the
case in the optically thick part of a GRB flow, inside its
photosphere.  The small fraction of the thermal energy that remains
when the flow passes through the photosphere then shows up as thermal
radiation emitted at the photosphere.  In the optically thin parts of
the flow, the radiative cooling times are typically quite short
compared with the expansion time scale as will be shown in
Sect.~\ref{sec:radloss}.  The medium stays cold, and all thermal
energy gets quickly converted into radiation.  We assume the radiation
processes to be similar to those invoked in the internal shock model,
so that the radiation is non-thermal.  Thus, we can only estimate the
spectrum of the (small) part that is emitted as thermal radiation at
the photosphere.  We are able to calculate the total amount of
non-thermal radiation, however, since this only depends on the rate
magnetic dissipation.

We make a two-zone approach with respect to the optical thickness to
simplify the treatment.  The flow is optically thick up to the
photosphere and matter and radiation are treated as one fluid in
permanent thermal equilibrium.  At the photosphere the internal energy
carried by the radiation decouples from the matter and escapes as
black-body radiation.  From the photosphere on, part of the Poynting
flux produces non-thermal radiation through the magnetic dissipation
process, while the rest still accelerates the flow.

\subsection{Evolution of the magnetic field}
\label{sec:evomag}
  
We model the evolution of the magnetic field $B$ in a Poynting flux
dominated outflow by a \emph{dissipation time scale} $\tau$ which
depends on the Alfv\'en speed and a typical length scale of the field
geometry considered.  The motivation and detailed derivation for this
approach is given in \citetalias{drenkhahn:02a}.  Since the
dissipation time scale $\tau$ depends on the local Alfv\'en speed in
the flow it is a function of the proper mass density $\rho$, the
proper internal energy $e$, the absolute value of the radial bulk
4-velocity $u$ and the magnetic field strength $B$: $\tau =
\tau(\rho,e,u,B)$.  We consider all thermodynamic quantities
$\rho,e,\ldots$ in the comoving frame while all other quantities
$B,u,\tau,\ldots$ refer to the lab frame, the frame in which the
central engine rests.  We use the notation $\Gamma = \sqrt{1+u^2}$ for
the bulk Lorentz factor of the outflow and $\beta = u/\Gamma$ for the
bulk velocity in units of the speed of light $c$.

The flow is assumed to be purely radial, and we are considering
distances from the central engine that are sufficiently far from the
Alfv\'en radius.  Thus any centrifugal acceleration of the flow has
already taken place, and the dominant field component is $B_\phi
\equiv B \gg B_r, B_\theta$.  Without internal dissipation, the
induction equation would thus yield $\partial_r \beta rB = 0$.  The
evolution equation for the magnetic field including dissipation is
equal to the induction equation for ideal MHD but with an additional
source term
\begin{equation}
  \label{eq:mfe}
  \partial_r \beta rB
  = - \frac{rB}{c\tau} \left(
    1 - \mu^2 \frac{(\beta rB)_0^2}{(\beta rB)^2}
  \right)
  \ .
\end{equation}
The index $0$ denotes quantities at some initial radius $r_0$ where
the dissipation starts.  The constant $\mu$ stands for the ratio
between the magnetic field component which cannot dissipate (described
by the ideal MHD induction equation) and the total field strength at
$r = r_0$.  Speaking in terms of Poynting flux this means that $\mu^2$
is (almost) equal to the Poynting flux fraction which does not
dissipate.  $\mu = 0$ corresponds to a complete decay of the magnetic
field while $\mu = 1$ means no dissipation at all.

We derived the functional form of the dissipation time scale $\tau$ in
\citetalias{drenkhahn:02a}:
\begin{equation}
  \label{eq:tau}
  \tau
  = \frac{2 \pi \Gamma^2}{\epsilon\Omega}
  \sqrt{1+u_\mathrm{A}^{-2}}
\end{equation}
where
\begin{equation}
 u_\mathrm{A}
 = \frac{B_\mathrm{co}}{\sqrt{4\pi w}}
 = \frac{B}{\sqrt{4\pi\Gamma^2 w}}
\end{equation}
is the Alfv\'en 4-velocity in the comoving frame and $w$ is the proper
enthalpy density.  In regions where the magnetic energy density
dominates over the matter energy density the Alfv\'en velocity is near
the speed of light so that the square root in (\ref{eq:tau}) is close
to 1.  This approximation was used in \citetalias{drenkhahn:02a} to
obtain analytical results.  In the present numerical study this
approximation is not made.

\subsection{Poynting flux and baryon loading}

A very important parameter of our model is the ratio between Poynting
flux and kinetic energy flux at the initial radius $r_0$ which we
denote by
\begin{equation}
  \label{eq:sigma0}
  \sigma_0  
  = \frac{L_\mathrm{pf,0}}{L_\mathrm{kin,0}}
  = \frac{\beta_0 r_0^2 B_0^2}{4\pi \Gamma_0\dot M c}
\end{equation}
where $\dot M$ is the mass flux per sterad.  The outflows of interest
for us are Poynting flux dominated so that $\sigma_0 \gg 1$.  The
initial Poynting flux ratio controls not only the initial velocity
but also the final velocity of the flow as explained below.

How the flow is accelerated from very low velocities near the source
is not possible to calculate with the approach presented since the
azimuthal velocity and the radial field components cannot be neglected
there.  We assume that magnetocentrifugal acceleration mechanisms work
there accelerating the flow up to the fast magnetosonic speed as it is
the case in stellar winds.  For relativistic Poynting flux dominated
winds we know that the typical length scale for this acceleration is
on the order of the Alfv\'en radius which is in size similar to the
light radius.  If the amount of initially injected thermal energy is
small (not much larger than the rest mass energy) it is converted
quickly into kinetic energy so that we can treat the flow to be cold
again at a few light radii.  In the cold limit the fast magnetosonic
speed equals the Alfv\'en speed.  The Alfv\'en 4-velocity is a
function of the initial Poynting flux ratio $\sigma_0$ only at
$r=r_0$: $u_\mathrm{A,0} = \sqrt{\sigma_0}$
\citepalias{drenkhahn:02a}.  We now take this value as initial
4-velocity for our numerical calculations $u_0 = \sqrt{\sigma_0}$
which start at $r_0\approx\mathrm{a~few}\times c/\Omega$.

In other GRB studies the baryon loading (mass flux) $\dot M$ and the
total energy flux $L$ determine the Lorentz factor by $\Gamma =
L/(\dot Mc^2)$.  It is often assumed that all the available energy is
converted into kinetic energy at first.  Under the same assumption we
showed in \citetalias{drenkhahn:02a} that the final Lorentz factor of
a dissipating Poynting flux outflow is $\Gamma_\infty =
\sqrt{1+\sigma_0^3} \approx \sigma_0^{3/2}$.  Hence, the baryon
loading $\dot M$ is determined from the total energy flux $L$ and the
Poynting flux ratio $\sigma_0$ by $\dot M = \sigma_0^{-3/2} L/c^2$.
It is a matter of taste weather one describes an outflow by $L,\dot M$
or $L,\sigma_0$ but we use the latter to keep the notation of
\citetalias{drenkhahn:02a}.

\subsection{The role of electric fields in the flow}
\label{sec:diffusivity}

In our present study we use the dynamic equations for ideal MHD flows
and have to make sure that the ideal MHD approximation is applicable.
In ideal MHD, the electric field vanishes in the frame moving with the
fluid.  But a non-vanishing comoving electric field must exist near
the reconnection centres for field annihilation to take place.  We
assume that the spacial regions occupied by a non-vanishing electric
field are small and only restricted to the reconnection centres.  Then
we can neglect its influence on the dynamic equations on larger scales
(see Sect.~\ref{sec:dyneq}).  But still, the field dissipation
produces an extra magnetic field gradient also on large scales
resulting in $\nabla^2\vec B\not=0$ in the lab frame.  We show in this
section how the large scale electric field can be estimated and that
it deviates only by a small component $\delta E$ from the electric
field of ideal MHD.  This component can be neglected in the numerical
calculations.

The following calculations are done in the lab frame for quantities
which vary only on large length scales.  We start with the induction
equation with non-vanishing conductivity $\sigma_c$:
\begin{equation}
  \label{eq:indeq}
  \partial_t\vec B 
  = \curl\left(\vec v\times\vec B\right)
  + \frac{c^2}{4\pi\sigma_\mathrm{c}} 
  \nabla^2 \vec B
  \ .
\end{equation}
Ohm's law reads
\begin{equation}
  \vec j
  = \sigma_\mathrm{c} \left(
    \vec E + \vec\beta\times\vec B
  \right)
  = \sigma_\mathrm{c} \cdot\delta\vec E
\end{equation}
so that one can substitute $\sigma_\mathrm{c}$ by $|\vec
j|/|\delta\vec E|$ in the stationary ($\partial_t\vec B=0$) induction
Eq.~(\ref{eq:indeq}):
\begin{equation}
  \label{eq:ind2}
  \curl\left(\vec v\times\vec B \right)
  = - \frac{c^2}{4\pi}
  \frac{|\delta\vec E|}{|\vec j|}
  \nabla^2 \vec B
  \ .
\end{equation}
Using the stationary form of Amp\`ere's law $\vec j = c\cdot\curl\vec
B/(4\pi)$ to eliminate $\vec j$ and solving (\ref{eq:ind2}) for
$|\delta\vec E|$ gives $|\delta\vec E|$ as function of $\vec v$ and
$\vec B$ only:
\begin{equation}
  |\delta\vec E|
  =
  \frac{|\curl\vec B| 
    \left|\curl\left(\vec v\times\vec B\right)\right|}
  {c |\nabla^2 \vec B|}
  \ .  
\end{equation}
In spherical coordinates and for a radial flow where $\vec v\perp \vec
B$ this reads
\begin{equation}
  \delta E
  = \left|
    \frac{(\partial_r rB)(\partial_r r\beta B)}{r\partial_r^2 (rB)}
  \right|
  \ .
\end{equation}
As fraction of the ideal MHD electric field $E_\mathrm{mhd}=\beta B$
that is
\begin{equation}
  \frac{\delta E}{\beta B}
  = \left|
    \frac{\partial_r\ln|r\beta B|}{\partial_r\ln|\partial_rrB|}
  \right|
  \ .
\end{equation}
This expression is a function of $r,\beta,B$ and can be calculated
numerically.  $\delta E/(\beta B)$ takes its maximum at radii where
the dissipation ceases where it can be of the order $0.1$ depending on
the chosen input parameters.  For the largest part of the flow $\delta
E/(\beta B)\ll 1$ so that the use of the ideal MHD equation for the
evolution of the field is justified.  The effect of small scale
reconnection processes is instead taken into account by the decay term
in (\ref{eq:mfe}).

\subsection{Dynamic equations}
\label{sec:dyneq}

The conservation equations for mass, energy, momentum together with
Eq.~(\ref{eq:mfe}) describing the evolution of the magnetic field
determine the proper mass density $\rho$, the proper internal energy
density (excluding the rest mass energy density) $e$, the radial
4-velocity $u$ and the magnetic field strength $B$ as functions of
radius.  In our model the mass, energy, and momentum equations read
\begin{equation}
  \label{eq:mass_cons}
  \partial_r r^2 \rho u = 0
  \ ,  
\end{equation}
\begin{equation}
  \label{eq:en_cons}
  \partial_r r^2 \left(
    w\Gamma u + \frac{\beta B^2}{4\pi}
  \right) 
  = 0
  \ ,
\end{equation}
\begin{equation}
  \label{eq:mom_cons}
  \partial_r r^2 \left(
    w u^2 + p
    + \left(1+\beta^2\right) \frac{B^2}{8\pi}
  \right) = 2rp
\end{equation}
\citep[cf.][]{koenigl:01,lyutikov:01}.  The variable $w$ denotes the
proper enthalpy density $w = \rho c^2 + e + p$ where $p$ is the
thermal pressure.  The thermodynamic quantities $\rho,e,p,w$ are
defined in the comoving frame.  We assume the gas (fully ionised
hydrogen) to be ideal with negligible heat conduction and an equation
of state $p = (\gamma-1) e$ (and $w=\gamma e$) where $\gamma$ is the
adiabatic index.

The continuity and energy Eqs.~(\ref{eq:mass_cons}),
(\ref{eq:en_cons}) are integrated to give the total mass loss per time
and per sterad
\begin{equation}
  \label{eq:Mdot}
  \dot M
  = r^2 u \rho c
\end{equation}
and the total luminosity per sterad
\begin{equation}
  \label{eq:L}
  L = \frac{w}{\rho c^2}\Gamma \dot M c^2
  + \beta c \frac{(rB)^2}{4\pi}
\end{equation}
where one identifies the kinetic energy flux per sterad
$L_\mathrm{mat}= w/(\rho c^2) \Gamma \dot M c^2$ and the Poynting
luminosity per sterad $L_\mathrm{pf}=\beta c (rB)^2/(4\pi)$.

\subsection{Below and beyond the photosphere}

As long as the medium is optically thick, matter and radiation can be
considered to be in thermal equilibrium.  In this case, $e$ includes
both the thermal particle energy and the radiation field energy.  The
pressure is dominated by the radiation so that the adiabatic index is
$\gamma=4/3$.  At the photosphere radius $r_\mathrm{ph}$, where the
outflowing material becomes optically thin, the radiation decouples
from the matter and escapes as black body radiation.  The pressure and
internal energy at radii $r > r_\mathrm{ph}$ is only provided by the
matter.

The transition from optically thick to optically thin conditions is
sharp, in practical flow models, and we simplify the computations here
by treating it as a discontinuity.  Its location is in principle found
by integrating the optical depth into the flow.  As in stellar
atmospheres and winds, a fair approximation for its location is the
point where the mean free path of a photon equals the density scale
height.  In our case, this is of the order of the distance $r$ from
the source.  At the photospheric temperatures encountered (a few keV)
the dominant opacity is electron scattering.  Taking into account the
Lorentz transformation of the mean free path to the rest frame, the
photospheric radius is thus given by $r_\mathrm{ph} =
\Gamma/(\kappa_\mathrm{Th}\rho)$ where $\kappa_\mathrm{Th}$ is the
Thomson scattering opacity and $\rho$ the density in the comoving
frame.

For $r > r_\mathrm{ph}$ the thermal pressure is only supplied by the
plasma and the adiabatic index $\gamma$ depends on the temperature.
For non-relativistic temperature $kT\ll m_\mathrm{e} c^2$ we have
$\gamma = 5/3$ but for hotter medium the electrons become relativistic
which lowers $\gamma$.  Since fast radiative cooling is a model
assumption the matter in the optically thin part stays cold enough so
that $\gamma = 5/3$ is valid there.  The validity is checked during
the numerical computations.

\subsection{Radiative loss}
\label{sec:radloss}

In the optically thin regime energy and momentum from the dissipating
magnetic field is transfered into radiative form.  The radiation
escapes and does not interact with the matter.  Let $\Lambda$ be the
emissivity of the medium in the comoving frame, that is the energy
which is radiated away per unit time and per unit volume.  If the
emission is isotropic in the comoving frame the energy and momentum
Eqs.~(\ref{eq:en_cons}), (\ref{eq:mom_cons}) including the radiative
loss terms are
\begin{equation}
  \label{eq:en_cons1}
  \partial_r r^2 \left(
    w \Gamma u + \frac{\beta B^2}{4\pi}
  \right) 
  = - r^2 \Gamma\frac{\Lambda}{c}
  \ ,
\end{equation}
\begin{equation}
  \label{eq:mom_cons1}
  \partial_r r^2 \left(
    w u^2 + p
    + \left(1+\beta^2\right) \frac{B^2}{8\pi}
  \right) 
  = 2rp - r^2 u\frac{\Lambda}{c}
\end{equation}
\citep{granot:01}.  

The importance of the cooling term depends on the cooling time
scale.  If it is short, the matter stays cold (gas pressure
negligible) during the dissipation process.  In this limit, all the
dissipated energy is locally radiated away.  Synchrotron emission is a
plausible fast cooling process.  It is particularly effective in our
model, because the magnetic field strengths are high in a Poynting
flux dominated outflow.  With (\ref{eq:B0}) one derives a typical
field strength
\begin{equation}
  B 
  \la 7\cdot10^7\,\mathrm{G} \cdot
  L_{50}^{1/2} r_{13}^{-1}
  \ .
\end{equation}
In the comoving frame this is
\begin{equation}
  B_\mathrm{co,3}
  \la 7\cdot10^2 \cdot
  L_{50}^{1/2} r_{13}^{-1} \Gamma_{2}^{-1}
  \ .
\end{equation}
The distance travelled by the medium in one cooling time in the lab
frame is $r_\mathrm{cool} = c t_\mathrm{syn} \Gamma$.  The synchrotron
cooling time scale in the comoving frame is $t_\mathrm{syn} =
6\,\mathrm{s} \cdot \Gamma_\mathrm{e,2}^{-1} B_\mathrm{co,3}^{-2}$
\citep{daigne:98} where $\Gamma_e = 100\,\Gamma_\mathrm{e,2}$ is the
Lorentz factor of the radiating fast electrons and $B_\mathrm{co} =
1000\,\mathrm{G}\cdot B_\mathrm{co,3}$ is the comoving magnetic field
strength.  In units of the expansion length scale of the flow, $r$,
the cooling time is
\begin{equation}
  \frac{r_\mathrm{cool}}{r}
  \approx 4\cdot 10^{-6} \cdot
  r_{13} L_{50}^{-1} \Gamma_2^3 \Gamma_\mathrm{e,2}^{-1}
  \ll 1
  \ .
\end{equation}
This shows that synchrotron cooling is fast for these fiducial
parameters.  Though the simplifying assumption of fast cooling in the
optically thin regime is thus justified, we have kept an ad hoc
cooling term in the calculations to ease the numerical treatment.

The form of this cooling term used is 
\begin{equation}
  \label{eq:Lambda}
  \Lambda
  = k \frac{ecu}{r}
\end{equation}
where $k$ is an adjustable cooling length parameter.  The cooling
length is the distance by which the matter travels outward while the
internal energy $e$ is lost.  We have used $k = 10^4$ so that the
cooling length is the distance $10^{-4}r$.  This is only a small
fraction of the expansion length scale $r$ and thus qualifies for the
description of a fast cooling flow.

Because of the fast cooling the temperature is always very low ($kT
\ll m_\mathrm{e}c^2$) and the equation of state of the gas is that of
a non-relativistic fully ionised gas, $\gamma=5/3$ for
$r>r_\mathrm{ph}$.

\subsection{Computational method}
\label{sec:compmeth}

We choose
\begin{equation}
  \label{eq:q}
  \vec q = \left(
    \begin{array}{c}
      r^2\rho c^2\\ r^2 e\\ u\\
      \frac{rB}{\sqrt{4\pi}}
    \end{array}
    \right)
\end{equation}
to be the vector of primitive variables.  There are no principle
reasons against taking e.g.\ ($\rho,e,u,B$) instead but the use of
(\ref{eq:q}) simplifies the following analytical expressions a bit.
The set of Eqs.~(\ref{eq:mass_cons}), (\ref{eq:en_cons1}),
(\ref{eq:mom_cons1}), (\ref{eq:mfe}) can be written in matrix form
\begin{equation}
  \label{eq:diffeq}
  \mathcal{A}\cdot \partial_r \vec q 
  = \vec s
\end{equation}
with the matrix
\begin{displaymath}
  \mathcal{A}
  =  \left(
    \begin{array}{cc}
      u       & 0\\
      u\Gamma & \gamma u\Gamma\\
      u^2     & \gamma\Gamma^2 - 1\\
      0 & 0
    \end{array}
  \right.
  \cdots  
\end{displaymath}
\begin{equation}
  \hfill
  \cdots
  \left.
    \begin{array}{cc}
      r^2 \rho c^2 & 0\\
      r^2 w\Gamma\left(1+\beta^2\right) 
      + \Gamma^{-3} \left(\frac{rB}{\sqrt{4\pi}}\right)^2 & 
      2 \beta \frac{rB}{\sqrt{4\pi}}\\
      2 r^2w u + \frac{\beta}{\Gamma^3} \frac{(rB)^2}{4\pi}
      & \left(1+\beta^2\right) \frac{rB}{\sqrt{4\pi}}\\
      \frac{rB}{\sqrt{4\pi}} \Gamma^{-3} & \beta
    \end{array}
  \right)
\end{equation}
and the source term vector
\begin{equation}
  \vec s
  =  \left(
    \begin{array}{c}
      0\\
      - r^2 \Gamma \frac{\gamma\Lambda}{c}\\
      2r(\gamma-1)e
      - r^2 u \frac{\gamma\Lambda}{c}\\
      - \frac{1}{c\tau}
      \frac{rB}{\sqrt{4\pi}}
      \left[
        1- \mu^2 \frac{(\beta r B)_0^2}{(\beta rB)^2}
      \right]
    \end{array}
  \right)
  \ .
\end{equation}
The elements of $\mathcal{A}$ and $\vec s$ are functions of $r, \vec
q$ and the constant model parameters $\sigma_0, \mu, \epsilon\Omega,
r_0$.  For the cooling term $\Lambda$ we use the expression
\begin{equation}
  \Lambda = \left\{
    \begin{array}{ll}
      0 &
      \mbox{for~} r \le r_\mathrm{ph}\\
      10^4 ecu/r &
      \mbox{for~} r > r_\mathrm{ph}
    \end{array}
  \right.
  \ ,
\end{equation}
the dissipation time scale (\ref{eq:tau}), and the adiabatic index
\begin{equation}
  \gamma = 
  \left\{
    \begin{array}{ll}
      4/3 &
      \mbox{for~} r \le r_\mathrm{ph} \\
      5/3 &
      \mbox{for~} r > r_\mathrm{ph}\\
    \end{array}  
  \right.
  \ .
\end{equation}

\subsubsection{Boundary conditions and solution process}

To initialise the solver at the initial radius $r_0$ we need to
determine the vector of primitive variables $\vec q_0 = \vec q(r_0)$.
The flow starts with Alfv\'en velocity $q_{0,3} = u_0 =
\sqrt{\sigma_0}$ and is cold $q_{0,2} = r_0^2 e_0 = 0$.  By solving
(\ref{eq:sigma0}), (\ref{eq:Mdot}), (\ref{eq:L}) for $\dot M$ we
obtain
\begin{equation}
  r^2 \rho uc 
  = \dot M
  = \frac{L}{c^2\left(\sigma_0+1\right)^{3/2}}
\end{equation}
and thus
\begin{equation} 
  q_{0,1}
  = r_0^2 \rho_0 c^2
  = \frac{L/c}{\sqrt{\sigma_0} \left(\sigma_0+1\right)^{3/2}}
  \ .
\end{equation}
These three equations can also be solved for $r_0B_0$ to give finally
\begin{equation}
  \label{eq:B0}
  q_{0,4}
  = \frac{r_0B_0}{4\pi}
  = \left(\frac{\sigma_0}{\sigma_0+1}\right)^{1/4}
  \sqrt{\frac{L}{c}}
  \ .
\end{equation}
The initial vector depends only on the initial radius, the initial
Poynting flux ratio and the total luminosity: $\vec q_0 = \vec
q(r_0;\sigma_0,L)$.  The value of $r_0$ is unimportant for the flow at
larger radii \citepalias{drenkhahn:02a} and we set $r_0 =
3\cdot10^7\,\mathrm{cm}$ for all our calculations without introducing
an additional restriction.  In total the model's input parameter space
is effectively made up of $\sigma_0, L, \mu, \epsilon\Omega$.

Eq.~(\ref{eq:diffeq}) is a ordinary differential equation and can be
solved numerically with common software packages.  The integration
proceeds stepwise from $r = r_0$ until the photosphere is reached,
where the mean free path for the photons is equal to the radius $r$.
The photosphere must be treated in a special way because it is a
discontinuity where the radiation decouples from the matter part.

\subsubsection{Transition at the photosphere of the flow}

At the photosphere, the equation of state changes from one dominated
by radiation to one dominated by the gas pressure.  To connect the
two, the radiation emitted at the photosphere has to be taken into
account.  The amount of energy involved can be substantial, and
appears as an (approximate) black body component in the GRB spectrum.
It depends on the temperature of the photosphere.

The temperature at the photosphere is $kT\ll m_e c^2$ for all used
parameter values so that pairs can be neglected.  The photosphere is
then simply determined by
\begin{equation}
  \left.
    \frac{\Gamma}{r\rho\kappa_\mathrm{Th}}
  \right|_{r_\mathrm{ph}}
  = 1
  \ .
\end{equation}

At the photosphere one has to subtract the energy and momentum which
is carried away by the decoupled radiation.  To calculate these
quantities one needs the temperature at the photosphere.

The dimensionless temperature $\theta=kT/(m_e c^2)$ in the optically
thick region is given by the solution of
\begin{equation}
  \label{eq:theta}
  e
  = 3 \frac{m_\mathrm{e}}{m_\mathrm{p}} \rho c^2 \theta
  + \frac{\pi^2}{15} \frac{m_\mathrm{e}c^2}{\cwl^3} \theta^4
\end{equation}
where $\cwl$ is the electron Compton wave length.  

At the photosphere we calculate the temperature $\theta_\mathrm{ph}$
and subtract the radiation energy density of a black body
\begin{equation}
  \label{eq:ebb}
  e_\mathrm{bb}
  = \frac{\pi^2}{15} \frac{m_\mathrm{e}c^2}{\cwl^3}
  \theta_\mathrm{ph}^4
\end{equation}
from the total energy density: $e \equiv e - e_\mathrm{bb}$.  The
integration proceeds with an adiabatic index of $\gamma = 5/3$.  The
temperature $\theta_\mathrm{ph}$ is the temperature of the emitted
black-body radiation which has a luminosity per sterad of
\begin{equation}
  \label{eq:Lbb}
  L_\mathrm{bb}
  = \left\{
    \begin{array}{ll}
      0 & \mbox{for~} r<r_\mathrm{ph}\\
      r_\mathrm{ph}^2 \frac{4}{3} e_\mathrm{bb}
      u_\mathrm{ph} \Gamma_\mathrm{ph} c &
      \mbox{for~} r\ge r_\mathrm{ph}
    \end{array}
  \right.
  \ .
\end{equation}

The integration continuous until the dissipation ceases.  There, the
luminosity of emitted non-thermal radiation is determined by
\begin{equation}
  \label{eq:Lnt}
  L_\mathrm{nt} 
  = L - L_\mathrm{pf} - L_\mathrm{mat} - L_\mathrm{bb}
  \ .
\end{equation}

\subsubsection{Modifications to the black-body radiation}
\label{sec:bbmod}

We have assumed in the above that the radiation emitted at the
photosphere is perfect black body radiation which means that the
photons are created at the photosphere.  But if scattering dominates
(scattering coefficient is of the same order or greater than
absorption coefficient) the photons from the photosphere are produced
at smaller radii (and at different temperatures) and undergo many
scatterings until they escape to infinity.
  
The number density of the photons is determined at the radius where
they are created.  The relevant creation processes are free-free and
Synchrotron emission.  Depending on the degree of Comptonisation the
emergent photon spectrum will be a modified black body or a Wien
spectrum \citep{rybicki:79}.  The typical photon energy is then given
to the photons by Comptonisation at some radius larger than the
creation radius but smaller than the photospheric radius.  We thus
have three characteristic radii for the radiation process: the photon
creation radius where the number density is determined, the
Comptonisation radius determining the typical photon energy and the
photosphere radius where the radiation decouples from the matter.
  
We can speculate about the spectrum of the emergent radiation if
scattering processes are included.  In thermal equilibrium the
radiative pressure-temperature relation reads $p\propto T^4$.  But if
the photon number does not depend on temperature any more the relation
changes to $p\propto T$ leading to an increase in temperature (similar
to the one seen at the photosphere in Fig~\ref{fig:sol-plot1}c).  The
photon production rate also rises until some equilibrium temperature
is reached.  Thus radiation and temperature cannot be determined
independently.  Compared to the black body temperatures at the
photosphere as used in the model presented the real temperatures may
thus be somewhat higher.  The number of photons might be less since
they originate from a smaller area.  Because the absorption and
emission processes are highly frequency dependent photons of different
energy might not be produced at the same radius and with the same bulk
Lorentz factor.  This could lead to a broader spectrum though
reprocessing by Comptonisation may also play a role.
  
A detailed modelling of the emergent radiation at the photosphere
requires a consistent but rather complicated extension of the model
presented here.  We have to assume that the simplified black body
treatment at least give reasonable estimates for the total energy
though the spectral shape might differ in reality.  The task to
determine the spectrum must be postponed to a more detailed
investigation in the future.

\subsubsection{Another transition radius}
\label{sec:srrad}

The rate of dissipation of the magnetic free energy starts out fast,
but as the field strength decreases it become slower than the
expansion time scale.  There is thus a characteristic radius, which we
call here the \emph{saturation} radius $r_\mathrm{sr}$, beyond which
magnetic dissipation effectively stops.  Since the acceleration of the
flow is intimately connected with the dissipation, this is also the
radius where the flow reaches its terminal speed.  In
\citetalias{drenkhahn:02a} we have shown that $r_\mathrm{sr}$ is of
the order
\begin{equation}
  \label{eq:rsr}
  r_\mathrm{sr} 
  = \frac{\pi c \Gamma_\infty^2}{3\epsilon\Omega}
  \ .
\end{equation}
while the terminal Lorentz factor $\Gamma_\infty$ is of the order
\begin{equation}
  \label{eq:Gam_inf}
  \Gamma_\infty 
  = \left(1-\mu^2\right) \sigma_0^{3/2}
  \ .
\end{equation}
A simple approximation for the dependence of the Lorentz factor on
distance then turns out to be
\begin{equation}
  \label{eq:Gam_ana}
  \Gamma 
  = \left\{
    \begin{array}{ll}
      \Gamma_\infty \left(r/r_\mathrm{sr}\right)^{1/3}
      & \mathrm{for~} r\le r_\mathrm{sr}\\
      \Gamma_\infty
      & \mathrm{for~} r> r_\mathrm{sr}
    \end{array}
  \right.
  \ .
\end{equation}

\subsubsection{Solution examples}
\label{sec:solex}

\begin{figure}
  \centerline{\includegraphics{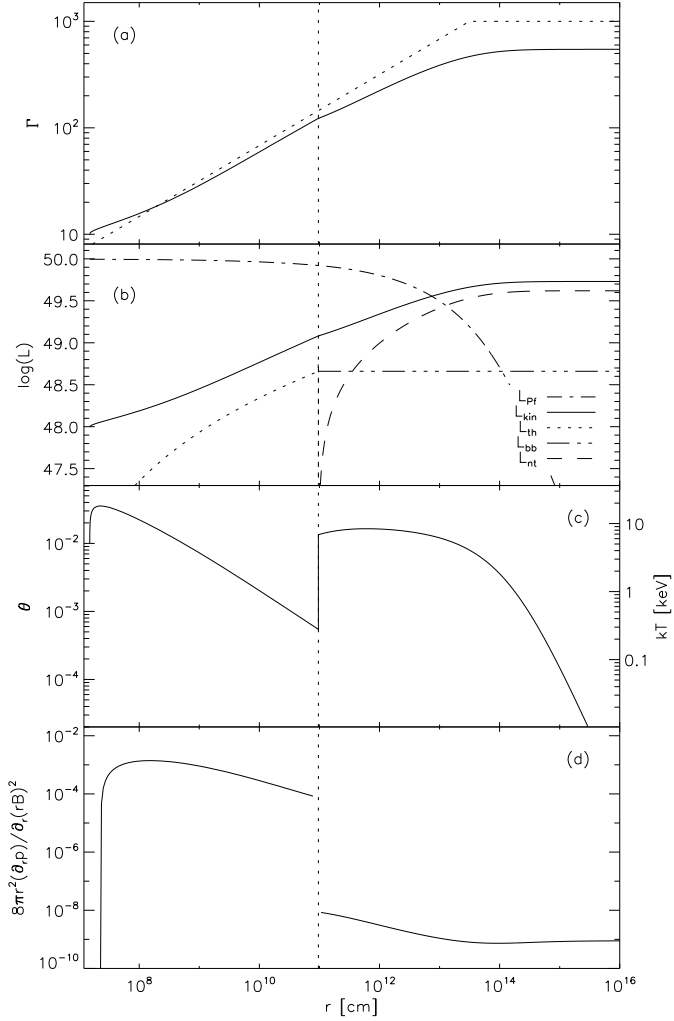}}
  \caption{Solution for 
    $L=10^{50}\,\mathrm{erg\,s^{-1}\,sterad^{-1}}$, $\sigma_0 = 100$
    (corresponding to an initial Lorentz factor of 10),
    $\epsilon\Omega = 10^3\,\mathrm{s^{-1}}$, $\mu = 0$.  The vertical
    dotted line indicates the photosphere radius.  (a) Lorentz factor
    where the dotted line represents the analytical approximation
    derived in \citetalias{drenkhahn:02a}.  (b) Various energy fluxes
    per sterad as labelled in the legend.  Indices denote the Poynting
    flux (pf), \emph{kin}etic, \emph{th}ermal, black-body (bb) and
    non-thermal (nt) components.  (c) Normalised temperature $\theta =
    kT/(m_\mathrm{e}c^2)$.  The matter is rather cold and pairs do not
    play a role since $\theta\ll1$.  The temperature jump at the
    photosphere is no real discontinuity but rather a rapid but steady
    change.  This is the result of the abrupt change in the equation
    of state from a radiation dominated to a matter dominated fluid.
    (d) Ratio between the thermal pressure gradient and the magnetic
    force density in the flow.  The acceleration is completely
    determined by the magnetic field at all radii.}
  \label{fig:sol-plot1}
\end{figure}

As an example, Fig.~\ref{fig:sol-plot1} shows the result of a
numerical integration.  The result can be compared with the analytical
approximation derived in \citetalias{drenkhahn:02a}.  The analytic
estimate gives a fair representation of the full results, though it
overestimates the terminal Lorentz factor somewhat.

The luminosity carried by the medium $L_\mathrm{mat} = w/(\rho c^2)
\Gamma\dot M c^2$ is made up of the kinetic $L_\mathrm{kin} =
\Gamma\dot M c^2$ and the thermal part $L_\mathrm{th} = (w/(\rho c^2)
- 1) \Gamma \dot M c^2$.  The fact that the fluid is dominated by the
pressure and energy density of the radiation in the optically thick
region can be seen at the photosphere.  Fig.~\ref{fig:sol-plot1}c
displays that a major part of the thermal energy flux $L_\mathrm{th}$
is made up from the radiation component which is released as black
body radiation at the photosphere.  This explains why the thermal
energy flux $L_\mathrm{th}$ in Fig.~\ref{fig:sol-plot1}b nearly
coincides with the black body luminosity $L_\mathrm{bb}$.

Outside the photosphere $\Gamma$ (and therefore $L_\mathrm{kin}$)
becomes smaller than the analytical estimate.  The non-thermal
radiation flux component $L_\mathrm{nt}$ rises quickly and the
dissipated energy is efficiently converted into radiation.  The
fractions of the total luminosity converted to kinetic, thermal, and
non-thermal energy in this example are 54\%, 5\%, and 41\%,
respectively.  This demonstrates the high efficiency of the magnetic
dissipation process in accelerating the flow and the production of
non-thermal radiation.

In \citetalias{drenkhahn:02a} we neglected the thermal pressure and
one could argue that the acceleration obtained in the present study
should be somewhat greater due to the additional effects of the finite
thermal pressure gradient of the matter and radiation component.  To
compare the magnetic and thermal contribution to the acceleration we
regard the momentum eq.~(\ref{eq:mom_cons}) which reads in
non-conservative form
\begin{equation}
  \frac{1}{r^2} \partial_r r^2 w u^2
  + \partial_r p
  = \frac{1}{c} \left(\vec j \times \vec B\right)_r
  = \frac{\partial_r (rB)^2}{8\pi r^2}
  \ .
\end{equation}
The ratio between the thermal pressure gradient $\partial_r p$ and the
magnetic force density $\left[\partial_r (rB)^2\right]/\left(8\pi
  r^2\right)$ is plotted in Fig.~\ref{fig:sol-plot1}d for our chosen
example parameter set.  At the photosphere the ratio decreases since
the radiation component drops out.  But even if radiation is included
the total thermal pressure gradient is not important compared to the
magnetic force density which determines the dynamics of the flow.
Thus the radiation pressure can be neglected above the photosphere
without influencing the dynamics of the flow.

\section{Shortest time scales}
\label{sec:time}

Though the model presented here is stationary and thus does not
describe the variability of GRBs, we can check whether the physical
model on which it is based is compatible with the observed millisecond
variations.  If the outflow contains inhomogeneities, say, regions
where the reconnection proceeds faster or unsteady the flow will show
more ore less bright patches.  In this section we show that the
emission time of these patches will be short enough for GRBs.

The time it takes for the magnetic field to dissipate sets the limit
for the shortest emission duration of a single patch in the flow.  The
time interval $\mathrm{d}t = \mathrm{d}r/c$ in which a flow element
moves outward by $\mathrm{d}r$ is Doppler boosted to the observed time
interval $\mathrm{d}t_\mathrm{obs} = \left(1-\beta\right) \mathrm{d}t
\approx \mathrm{d}r/(2c\Gamma^2)$.  Let us assume that the more
inhomogeneous flow can still be described approximately by the
stationary solution.  Using (\ref{eq:Gam_ana}) and integrating the
observed time from $r = 0 \ldots r_\mathrm{sr}$ gives
\begin{equation}
  t_\mathrm{obs}
  = \frac{\pi}{2\epsilon\Omega}
  = 1.6 \cdot 10^{-3}\,\mathrm{s}
  \cdot \left(\epsilon_{-1} \Omega_4\right)^{-1}
  \ . 
\end{equation}
If the flow consists of shells with differently strong or fast
reconnection/emission it can produce observable variations on time
scales of the order of milliseconds.

To see if small patches can contribute a significant variability, we
note that their lateral size (perpendicular to the flow) is likely to
be of the same order as $\lambda = 2\pi c/\Omega$.  This is the
typical size of the small scale field inhomogeneities and the
reconnection centres.  Due to the short cooling time the emitted
radiation originates from close to the reconnection centres so that
$\lambda$ is also the lateral size of the bright patches.  The overall
expansion in the flow does not change the size of the patches if the
expansion speed in the comoving frame is always small compared to the
Alfv\'en speed $\approx c$ which is the typical speed with which the
magnetic field can reorganise itself.  This is the case for
$\lambda<r/\Gamma$ being always true at the radii of interest.  Thus
these patches will stay of the same size $\lambda=$const.\ and do not
scale with radius.  $\lambda$ is quite small compared with the radius
of the photosphere.  The solid angle from which the observed radiation
is emitted is of the order $1/\Gamma^2$.  Hence there are $n =
(r/\lambda\Gamma)^2 = 3\cdot10^{11} \cdot \Omega_4^2 \Gamma_2^{-2}
r_{13}^2$ reconnection patches contributing at any moment to the
observed radiation, and the maximum variability amplitude expected is
thus $\sqrt{n}/n \approx 2\cdot 10^{-6}$.  The observed time scales
are not produced locally in the flow but must be due to a variability
of the central engine.

\section{Parameter study}

The parameters of the model are the initial Poynting flux to kinetic
energy flux ratio $\sigma_0$, the total luminosity per sterad $L$, the
fraction of non-dissipatable magnetic field $\mu$, and the measure for
the reconnection rate $\epsilon\Omega$.  In this section we explore the
dependence of the solutions on these parameters, by plotting values of
the physical quantities at the photosphere and their asymptotic values
at large distances.

\subsection{The photosphere}

\begin{figure}
  \centerline{\includegraphics{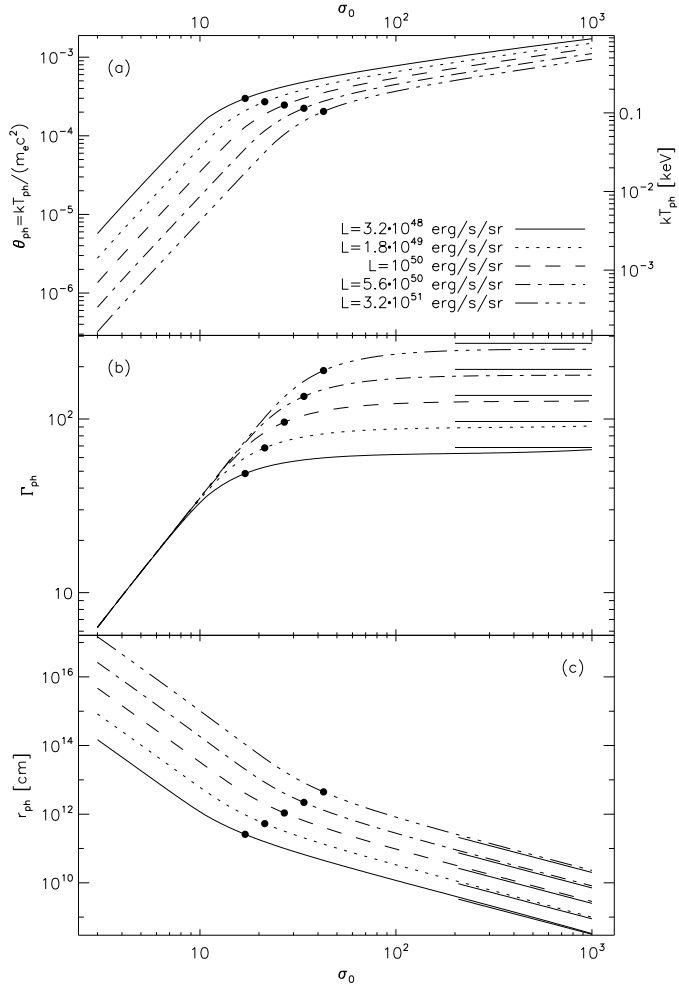}}
  \caption{Some physical quantities at the photosphere of the flow
    as function of the initial Poynting flux ratio $\sigma_0$ for
    different values of the total luminosity per sterad.  The values
    $\epsilon\Omega = 10^{-3}\,\mathrm{s^{-1}}$, $\mu = 0$ where used.
    (a) Temperature in the comoving frame.  (b) Lorentz factor at the
    photosphere.  (c) Radius of the photosphere.  Straight lines in
    panel (b)+(c) at $200 < \sigma_0 < 1000$ indicate the analytical
    solutions from (\ref{eq:rph}), (\ref{eq:uph}).  Thick dots
    correspond to $\sigma_0 = \sigma_{0,\mathrm{br}}$ defined in
    (\ref{eq:sigbr}).}
  \label{fig:x_over_sigma-ph}
\end{figure}

Figure~\ref{fig:x_over_sigma-ph} shows the temperature, Lorentz factor
and radius of the flow at the photosphere as functions of the initial
Poynting flux ration $\sigma_0$.  In all plots of the three quantities
one can identify a break at a certain value of $\sigma_0$.  This break
can be understood in terms of the saturation radius $r_\mathrm{sr}$
mentioned in Sect.~\ref{sec:solex}.  The reconnection yields only
little energy beyond $r_\mathrm{sr}$ since the largest part of the
free magnetic energy is already used up before.  The location of this
radius relative to the photosphere determines the basic properties of
the results.  In order to interpret the numerical results, we recall
here some the results of the analytic model derived in
\citetalias{drenkhahn:02a}.  In this model the radius of the
photosphere $r_\mathrm{ph}$ and the Lorentz factor at the photosphere
are given by
\begin{eqnarray}
  r_\mathrm{ph}
  &=& 1.05\cdot10^{11}\,\mathrm{cm}\nonumber\\
  && \cdot \left[
    \epsilon_{-1} \Omega_4 \left(\frac{1-\mu^2}{0.5}\right)
  \right]^{-2/5}
  L_{50}^{3/5}
  \sigma_{0,2}^{-3/2}
  \ ,
  \label{eq:rph}\\
  \Gamma_\mathrm{ph}
  &\approx&
  u_\mathrm{ph}
  = 119\cdot \left[
    \epsilon_{-1} \Omega_4
    \left(\frac{1-\mu^2}{0.5}\right) L_{50}
  \right]^{1/5} 
  \ .
  \label{eq:uph}
\end{eqnarray}
By equating $r_\mathrm{ph}$ and $r_\mathrm{sr}$ of (\ref{eq:rph}),
(\ref{eq:rsr}), we find the value of $\sigma_0$ where the break in the
parameter dependences occurs:
\begin{equation}
  \label{eq:sigbr}
  \sigma_{0,\mathrm{br}}
  =
  39\cdot
  \left(
    \epsilon_{-1} \Omega_4 L_{50}
  \right)^{2/15}
  \left(\frac{1-\mu^2}{0.5}\right)^{-8/15}
  \ .
\end{equation}
At this value of the baryon loading, the dissipation of magnetic
energy ceases around the photosphere.  For $\sigma_0 <
\sigma_{0,\mathrm{br}}$ dissipation occurs mostly inside the
photosphere, and the radiation is dominated by a black body component.

We note that the analytic expression (\ref{eq:rph}) for
$r_\mathrm{ph}$ has been derived under the assumption that most of the
dissipation occurs outside the photosphere.  It is still accurate
enough, however, for the estimate (\ref{eq:sigbr}) that we use to
interpret the numerical results.  The asymptotic validity of the
analytic value (\ref{eq:uph}) of the terminal Lorentz factor
$\Gamma_\infty$ for large $\sigma_0$ is shown in
Fig.~\ref{fig:x_over_sigma-ph}b.

\subsection{Limits on thermal and non-thermal radiation}

The model yields the luminosity per sterad for both the black-body
radiation from the photosphere and the non-thermal radiation.  In this
section we investigate how these radiation components behave as function
of the model parameters and what observed temperatures are expected
for the black-body component.

\begin{figure}
  \centerline{\includegraphics{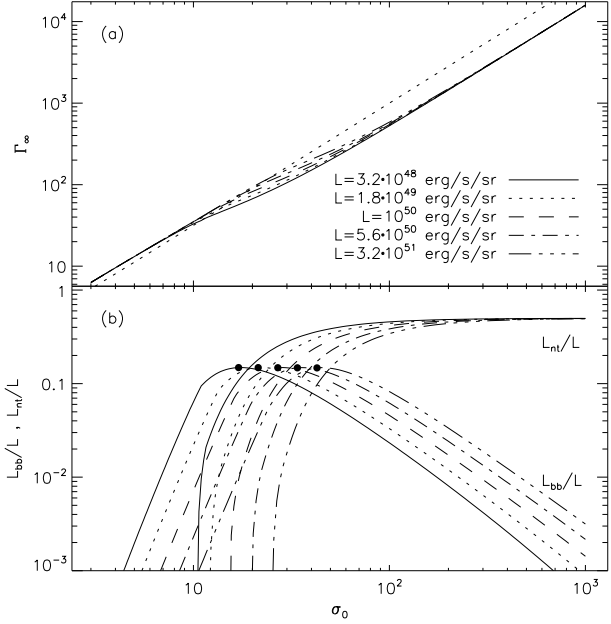}}
  \caption{Terminal Lorentz factor and radiation flux ratios
    as function of the initial Poynting flux ratio $\sigma_0$ for
    different values of the total luminosity per sterad.  The values
    $\epsilon\Omega = 10^{-3}\,\mathrm{s^{-1}}$ and $\mu = 0$ where
    used.  (a) Terminal Lorentz factor $\Gamma_\infty$.  The dotted
    line correspond to the analytical estimates from
    \citetalias{drenkhahn:02a} where $\sigma_0 \gg 1$ and no radiative
    losses were assumed.  (b) Ratio between black-body and total
    luminosity $L_\mathrm{bb}/L$ and ratio between non-thermal and
    total luminosity $L_\mathrm{nt}/L$.  At the location of the thick
    dots, the model parameters are such that the magnetic dissipation
    ceases to be effective near the photosphere [cf.
    (\ref{eq:sigbr})].}
  \label{fig:x_over_sigma}
\end{figure}
\begin{figure}
  \centerline{\includegraphics{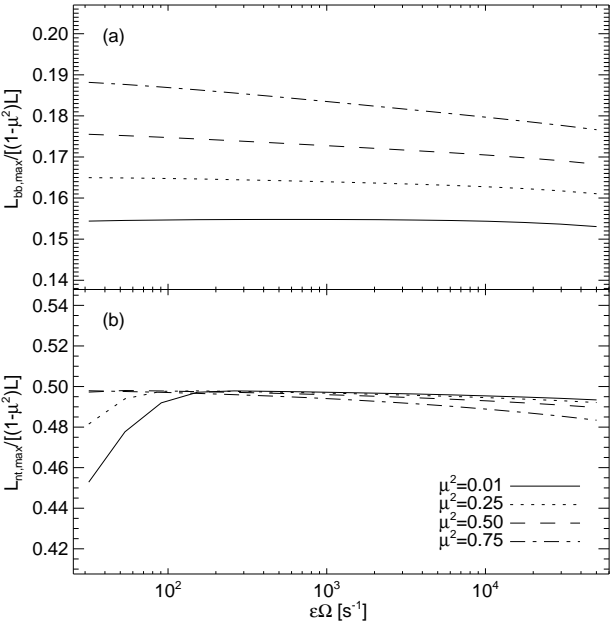}}
  \caption{Maximal thermal and non-thermal emission as function of
    $\epsilon\Omega$ (a measure for the dissipation rate).  (a) Ratio
    of maximal thermal emission to dissipatable luminosity
    $L_\mathrm{bb,max}/[(1-\mu^2)L]$.  The used values for
    $L_\mathrm{bb,max}$ correspond to the thick dots in
    Fig.~\protect\ref{fig:x_over_sigma}b.  (b) Ratio of maximal
    non-thermal emission to dissipatable luminosity
    $L_\mathrm{nt,max}/[(1-\mu^2)L]$ determined at $\sigma_0 = 10^3$.}
  \label{fig:maxL}
\end{figure}

Figure~\ref{fig:x_over_sigma}b displays the luminosities of the
thermal and non-thermal radiation components $L_\mathrm{bb}$ and
$L_\mathrm{nt}$ as fraction of the total luminosity $L$.  At very low
$\sigma_0$-values both components are very small.  In that case the
energy is released far below the photosphere and is converted into
kinetic energy.  The same happens in `dirty fireball' models where the
central engine injects thermal energy into the matter near the source
\citep{shemi:90,paczynski:90}.  This also leads to an almost complete
conversion into kinetic energy.

The black-body radiation shows its maximum if the dissipation ceases
right at the photosphere so that $r_\mathrm{ph}=r_\mathrm{sr}$.  This
corresponds to $\sigma_0=\sigma_{0,\mathrm{br}}$ from
(\ref{eq:sigbr}).  This analytical estimate for
$\sigma_{0,\mathrm{br}}$ might be not coincide exactly with the
maximum of the numerically obtained $L_\mathrm{bb}/L$-curves in
Fig.~\ref{fig:x_over_sigma}b since some simplifications were used in
the derivation of (\ref{eq:sigbr}).  Though, we take
$L_\mathrm{bb,max} = L_\mathrm{bb}(\sigma_\mathrm{0,br})$ to be the
maximal black-body luminosity to simplify the treatment in the
following.

From the graphs in Fig.~\ref{fig:x_over_sigma}b one might guess that
the maximal value of $L_\mathrm{bb}/L$ does not depend on $L$.
Indeed, it turns out that the maximal fraction of the black-body
luminosity to the \emph{dissipatable} luminosity is almost a constant.
In the Poynting flux dominated wind the initial Poynting flux
luminosity is almost equal to the total luminosity
$L_\mathrm{pf,0}\approx L$.  The fraction of Poynting flux which
cannot dissipate by reconnection was parameterised by $\mu^2$ so that
dissipatable luminosity is $(1-\mu^2)L_\mathrm{pf,0}\approx
(1-\mu^2)L$.  If we plot $L_\mathrm{bb,max}/[(1-\mu^2)L]$ as a
function of the two other model parameters $\mu,\epsilon\Omega$ we see
that its value is around $0.17\pm0.03$ as displayed in
Figure~\ref{fig:maxL}a.  Thus, the energy in black-body radiation is
always less than 20\% of the total releasable magnetic energy.

Figure~\ref{fig:x_over_sigma}b shows that the fraction of the total
luminosity emitted as non-thermal radiation has a maximum value for
large $\sigma_0$ of about 50\%, independent of the luminosity itself.
In this limit almost all the dissipation takes place outside the
photosphere and $L_\mathrm{bb}/L$ is negligible.

Figure~\ref{fig:maxL}b shows that $L_\mathrm{nt,max}/[(1-\mu^2)L]$ is
always very close to $0.5$.  The maximal radiation efficiency occurs
in the extreme Poynting flux dominated limit.  A fast radiation
mechanism converts half of the free magnetic energy into non-thermal
radiation.

The dissipatable energy flux $(1-\mu^2)L$, not the total Poynting flux
$L_\mathrm{pf} \approx L$ in general, is the energy reservoir from
which the radiation and kinetic energy is fed.  Nevertheless, one
needs to know the total Poynting flux in order to determine the
absolute magnetic field strength in the medium to investigate the
physical emission process.  Since this is not needed here we could
restrict our study to the case $\mu=0$ in the largest part of this
paper.  For any combination of $\mu,L$ one finds $\mu'=0$,
$L'=(1-\mu^2)L$ which yield an outflow of equal dissipatable Poynting
flux and thus equal emission and kinetic energy.  Therefore, the
setting $\mu=0$ has not introduced an additional limitation in the
generality of our investigation.

\subsection{Limits on the terminal Lorentz factor}
\label{sec:Gamma_inf}

In \citetalias{drenkhahn:02a} we derived the terminal Lorentz factor
for a complete conversion of Poynting flux into kinetic energy flux.
For a Poynting flux dominated flow this terminal Lorentz factor is
$\Gamma_\infty = (1-\mu^2)\sigma_0^{3/2}$.  Since we found that the
radiative losses can be as large as 50\% of the total luminosity we
can now determine upper and lower limits for the terminal Lorentz
factor:
\begin{equation}
  \frac{1}{2} \left(1-\mu^2\right) \sigma_0^{3/2}
  < \Gamma_\infty
  < \left(1-\mu^2\right) \sigma_0^{3/2}
  \ .
\end{equation}

\subsection{Possible variability}
\label{sec:var}

The stationary treatment in our study does not yield any variability
by definition.  But we can speculate about the outcome of
quasi-stationary changes in one or more model parameters.  The
non-thermal luminosity $L_\mathrm{nt}$ depends most strongest on
$\sigma_0$ at moderate values.  This is seen in
Fig.~\ref{fig:x_over_sigma}b where the $L_\mathrm{nt}/L$ graphs rise
quickly to the maximal values in an values around $20 \la \sigma_0 \la
100$.  A variation of $\sigma_0$ in this moderately large interval
might produce some kind of on-off behaviour of the non-thermal
luminosity and the large variability in GRB light curves.  Extending
the model to include time dependence remains an interesting
investigation for the future.

\subsection{Observable quantities}
\label{sec:obs}

The results show that a thermal component in the emission is expected
at higher baryon loading values.  In a limited range of baryon loading
($10\la\sigma_0\la70$), the model predicts a GRB with a significant
thermal component.  This property can be used observationally as a
test of the model, or as a diagnostic of the GRB outflow.

\begin{figure}
  \centerline{\includegraphics{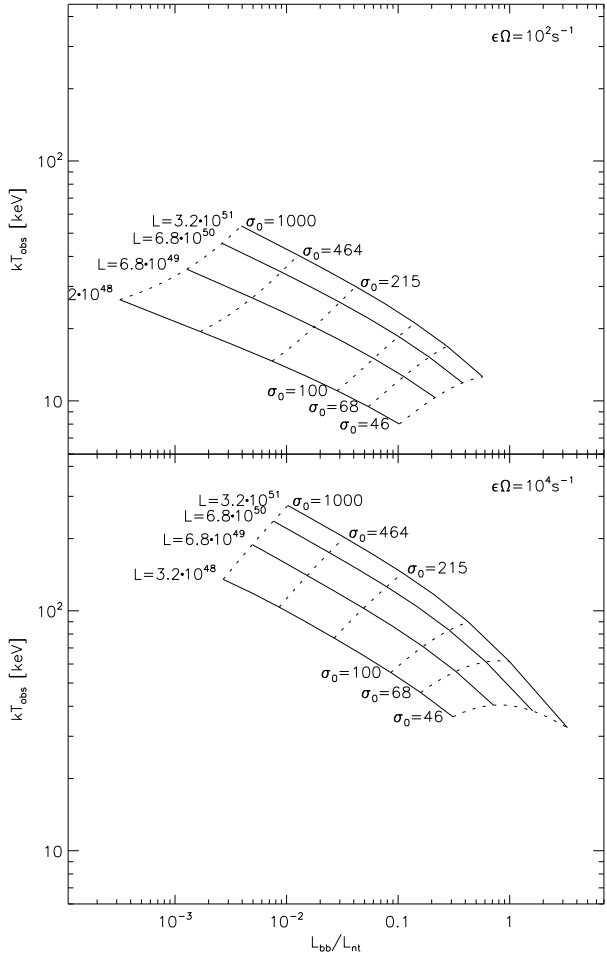}}
  \caption{Thermal to non-thermal flux ratio
    $L_\mathrm{bb}/L_\mathrm{nt}$ and redshift corrected temperature
    of thermal radiation for various model parameters.  Calculations
    for each panel are done for fixed values of $\mu = 0$ and
    $\epsilon\Omega$ as indicated.  Solid lines correspond to
    parameter sets with equal total luminosity $L$ while dotted lines
    correspond to equal initial Poynting flux ratio $\sigma_0$.}
  \label{fig:kTLbbLnt}
\end{figure}

Our model GRBs can be represented in a diagram of showing the ratio
$L_\mathrm{bb}/L_\mathrm{nt}$ as a function of the black body
temperature.  This is shown in Fig.~\ref{fig:kTLbbLnt}.  The
photospheric (black body) temperature $T_\mathrm{obs}$ shown is the
value as observed in the rest frame of the GRB host.  It is related to
the temperature in a comoving frame $T_\mathrm{bb}$ by $T_\mathrm{obs}
= \Gamma_\mathrm{ph} T_\mathrm{bb}$.

The single panels of Fig.~\ref{fig:kTLbbLnt} show lines for which 3 of
the 4 model parameters are fixed while $L$ or $\sigma_0$ is varied.
An increase in the initial Poynting flux ratio $\sigma_0$ increases
the temperature but decreases the black-body component due to a
smaller photosphere radius (see Fig.~\ref{fig:x_over_sigma-ph}c).  An
increase of the total luminosity $L$ results in a larger photosphere
temperature and a larger black-body component though the dependence of
$L_\mathrm{bb}/L_\mathrm{nt}$ and $kT_\mathrm{obs}$ on $L$ is much
weaker than on $\sigma_0$.

From Fig.~\ref{fig:kTLbbLnt} one finds that only low Poynting flux
ratios $\sigma_0\la 200$ lead to a significant fraction of thermal
radiation $L_\mathrm{bb}/L_\mathrm{nt}\ga 0.1$.  The predicted black
body temperatures can range all the way from about 5 to 100\,keV.
These are the temperatures as observed in the frame of the GRB host,
so a redshift has to be known in order to use these predictions
diagnostically.

Thermal components have indeed been observed in GRB spectra.
\citet{preece:02} reports on a thermal component in the spectra of
GRB\,970111 within the first $\approx10\,\mathrm{s}$ after the
trigger.  In this time interval the temperature of the black body
component was observed to vary between 45 and 75\,keV, and the ratio
of thermal to non-thermal flux was of the order unity.  After this
initial phase the non-thermal component started to dominate.  In terms
of our model, these observations indicate that this GRB started with a
moderate baryon loading, which then decreased in the course of the
burst.  Since a redshift of this burst has not been determined, a more
detailed comparison with the model can not be made.

\subsection{Connection with X-ray flashes}
\label{sec:xray}

X-ray flashes are fast X-ray transients which are not detected in the
$\gamma$-ray band $40$--$700\,\mathrm{keV}$ of BeppoSAX
\citep{heise:01,heise:02}.  In our model this spectral characteristic
can be explained by an outflow of low $\sigma_0$ where the thermal
radiation dominates over the non-thermal component.  The relevant
region in Fig.~\ref{fig:kTLbbLnt} would be at
$L_\mathrm{bb}/L_\mathrm{nt} \ga 1$.  The temperatures predicted by
the model for this range of black body luminosity is $\la
30\,\mathrm{keV}$, which is quite compatible with the observations.

\section{Summary and discussion}

A magnetised and rotating central engine of a GRB produces a Poynting
flux outflow.  Any non-axisymmetry of the magnetic field leads to
small scale (wave-like) variations in the electromagnetic field
carrying energy outward.  We have assumed there that these small scale
irregularities are subject to rapid reconnection, governed by the
Alfv\'en speed, as observed in other astrophysical settings and in
numerical simulations.  Thus, the magnetic field can rearrange itself
to a energetically favourable configuration and releases its free
energy stored in the small scale field variations.

The release of free magnetic energy proceeds with a rate determined by
the length scale of the field variation and the local Alfv\'en speed
of the plasma.  The magnetic field acts as energy reservoir carried
with the matter which transfers its energy to the matter continously.
The decay of the magnetic field at the same time causes an outward
gradient of the magnetic pressure.  This causes a significant part of
the Poynting flux to be converted into kinetic energy.  The other part
of the free energy is converted into heat.  In the optically thick
region of the flow a thermal energy gradient promotes adiabatic
expansion and the conversion of thermal energy to kinetic energy.
Thus, at small radii in the optically thick region, almost all of the
dissipated magnetic energy gets converted into kinetic energy.

When the flow becomes optically thin at the photosphere the thermal
radiation energy escapes to infinity.  This radiation resembles a
black body spectrum if the photons are created near the photosphere.
But if scattering dominates the spectral shape will differ and might
look like e.g.\ a modified black body or a Wien spectrum
\citep{rybicki:79}.  A much more detailed treatment is necessary to
determine the radius at which the radiation is produced and what
spectral shape it has.  At this stage we are only interested in the
total energetics and the characteristic radiation temperature and we
assume that the simplified black body treatment give sufficiently
precise estimates.

To compute the radiation spectrum from the optically thin region more
detailed physics is needed which is beyond the scope of this paper.
We have instead assumed that the reconnection process under optically
thin conditions maintains a significant population of energetic
electrons, which then radiate synchrotron radiation in much the same
way as in the standard internal shock model.  An advantage of our
model is that the magnetic field needed for the synchrotron radiation
is a natural part of the flow model itself (see also
\citetalias{spruit:01}).  The central difference of the model
presented to standard internal shock models for GRBs is that radiation
stems from the local dissipation of magnetic energy and not from shock
conversion of kinetic energy.  Therefore, one does not need an
extremely variable central engine to obtain an acceptable radiation
efficiency \citep{beloborodov:00,kobayashi:01}.

The continuous character of the energy release leads to a slower
acceleration of the flow compared to the classical fireball scenario.
In a fireball the energy is injected abruptly as thermal energy.  This
leads to an rapid acceleration where the Lorentz factor is linear to
the source distance $\Gamma\propto r$.  In our model the release of
magnetic energy leads to $\Gamma\propto r^{1/3}$ in the optically
thick region.

An important model parameter is the ratio between Poynting flux and
kinetic energy flux $\sigma_0$ at some initial radius $r_0$.  This
parameter controls the baryon loading parameter in a sense that high
values correspond to a low baryon loading.  The value of $\sigma_0$
decides how much of the Poynting flux energy gets converted into
kinetic energy, black-body radiation and non-thermal radiation.  The
three other parameters of the model are the total luminosity per
sterad, the fraction of dissipatable Poynting flux and the
reconnection speed.  There are three intervals of $\sigma_0$ values in
which the characteristics of the flow is significantly different.  At
very low values ($\sigma_0\la 10$) most of the energy gets converted
into kinetic energy.  The magnetic energy gets released in the
optically thick part.  This case is similar to dirty fireball models
in which the central engine injects thermal energy into the matter
near the central engine \citep{shemi:90,paczynski:90}.  The matter is
already cold when it reaches the photosphere at large radii and there
is no more free magnetic energy available to power the non-thermal
radiation.  The burst energy can only power an afterglow by an
external shock.

An intermediate Poynting flux ratio of $\sigma_0\approx 100$ causes
the release of a considerable amount of energy near the photosphere
and the thermal emission is non-negligible.  The black-body component
becomes maximal if the radius where the dissipation ceases coincides
with the photosphere.  Then, $\approx 17\%$ of the dissipatable
magnetic energy gets converted into black-body radiation while the
rest ends up in kinetic energy.

At very high $\sigma_0\la 300$ values the radius of the photosphere is
small and almost all of the dissipation takes place in the optically
thin region.  The dissipated energy gets equally distributed among the
non-thermal radiation and the kinetic luminosity of the flow.  At low
baryon loading, and for a purely non-axisymmetric magnetic field,
almost exactly 50\% of the the Poynting is converted into kinetic
energy and 50\% into non-thermal radiation.  If a substantial part of
the magnetic field is axisymmetric, the Poynting flux associated with
it does not dissipate, and instead is expected to show up as afterglow
emission.

Assuming that regular GRBs have large $\sigma_0$ this finding predicts
that the energy of the afterglow (fed by the kinetic energy of the
flow) is comparable to the energy of the the prompt emission.  Beaming
effects change this picture if the outflow consists of a sufficiently
narrow jet.  The energy in the afterglow will be weaker because after
the flow has decelerated the radiation is spread over a larger solid
angle compared to the highly beamed initial radiation \citep[cf.\ the
light curve break discussion in e.g.][]{ghisellini:01}.  Therfore the
prompt emission might be more luminous than the afterglow luminosity
if the jets points towards us.

Besides the luminosity of the black-body radiation the model yields
the temperature of this radiation.  We find that the unredshifted
observable temperature is $5\,\mathrm{keV} \la kT_\mathrm{obs} \la
100\,\mathrm{keV}$ for our fiducial GRB/X-ray burst parameters.  The
model produces a rather constant Lorentz factor at the photosphere so
that large variations of model parameters result in only a small
temperature spread.  We cannot make a clear statement about the
contribution of the non-thermal component to emission in the quoted
energy range because we do not know the radiation mechanism.  If the
thermal component at its maximum is of the same order or greater than
the non-thermal component a feature should be present in the spectrum.
In fact, there exist observations of excess emission in the low energy
range ($\approx 1$--$5\,\mathrm{keV}$) for some GRBs
\citep{strohmayer:98,preece:96}.  Recent investigations by
\citet{preece:02} show clearly a strong thermal component in the
during the first 10\,s of GRB\,970111 with temperatures of
45--75\,keV.  These numbers are in agreement with some of our fiducial
GRB model parameter values.  Observations of this kind will enable us
to determine the model parameters and even there time-dependence
during the burst.

For initial ratios between Poynting flux and kinetic energy flux
$\sigma_0\approx40$ the black-body radiation component dominates over
the non-thermal component.  The radiation efficiency is lower than in
the $\sigma_0\ga300$ case because only $\la17\%$ of the total
luminosity can be converted into radiation.  We speculated in
\citetalias{drenkhahn:02a} about the possibility that these
low-$\sigma_0$ outflows could be identified with X-ray flashes
observed by BeppoSAX \citep{heise:01,heise:02}.  Because no
dissipation takes place outside of the photosphere there is no
non-thermal emission in the $\gamma$-ray range $>40\,\mathrm{keV}$.
The present study showed that the thermal emission has an
(unredshifted) observable temperature of $\la30\,\mathrm{keV}$ which
agrees with the observations of X-ray flashes.  \citet{heise:01}
speculated that X-ray flashes are bursts with high mass loading.  This
is also true in our model since high mass loading corresponds to low
$\sigma_0$ values.

The hypothesis that the thermal radiation from a low-$\sigma_0$
outflow produces a X-ray flash may be checked by future observation.
The model predicts a lower radiation efficiency of $\la17\%$ so that
most of the dissipated energy in the outflow goes into kinetic form.
It will be converted into radiation in the external shock of the
afterglow.  The afterglow of an X-ray flash should be more luminous
than the prompt emission if jet-effects do not interfere too much.

A stationary approximation for the flow is used in this paper.  If the
central engine operates intermittently internal shocks could occur in
the magnetised outflow.  One could also imagine that not the total
luminosity changes with time but that the other wind parameters like
the mass loading and with it $\sigma_0$ varies.  $\sigma_0$ has a
strong influence on the non-thermal luminosity $L_\mathrm{nt}$ and the
black-body luminosity $L_\mathrm{bb}$.  A time-varying $\sigma_0$
around intermediate values leads certainly to large modulations in the
non-thermal light curve.  The physical model should be extended to
include time dependent model parameters to investigate their effect on
the light curve.

The minimal observed variability of GRBs is around a millisecond.  Any
process producing the emission must therfore be fast enough to account
for this limit.  The stationary model predicts that the reconnection
in the comoving frame lasts for approximately 1 millisecond.  The
effects for Doppler shift and relativistic time dilation almost cancel
so that one observes almost the time in the comoving frame.  The
millisecond variability is therfore compatible with the reconnection
model.  If the reconnection in the flow is not smoothly distributed
but patchy we would expect to see a peak for the emission coming from
one of the patches where reconnection takes place.  Our model is
nevertheless applicable because we only need the overall reconnection
rate, the average over small length scales, which is responsible for
the global flow dynamics.

\bibliographystyle{aa}
\small
\bibliography{h3507}
\end{document}